\documentclass[prx, twocolumn,  superscriptaddress]{revtex4-2}
\usepackage{setspace}
\usepackage[table]{xcolor}
\usepackage{tabularx}
\usepackage[colorlinks, citecolor=blue, linkcolor=blue, urlcolor=blue, linktocpage]{hyperref}

\usepackage[utf8]{inputenc}
\usepackage{graphicx}
\usepackage{amsmath}
\usepackage[]{natbib}

\usepackage{graphicx}
\usepackage{xcolor}
\setcitestyle{numbers}

\usepackage{eurosym}
\usepackage{tikz}
\usepackage{adjustbox}
\usepackage{amsthm}

\usepackage{enumitem}
\setitemize{noitemsep,topsep=0pt,parsep=0pt,partopsep=0pt,leftmargin=*}
\usepackage{amssymb}

\usepackage{mdframed}
\usepackage{enumitem}

\begin{document}
\title{Solving non-native combinatorial optimization problems using hybrid quantum-classical algorithms}

\author{Jonathan Wurtz}
\affiliation{QuEra Computing Inc., 1284 Soldiers Field Road, Boston, MA, 02135, USA}

\author{Stefan H. Sack}
\affiliation{Institute of Science and Technology Austria (ISTA), Am Campus 1, 3400 Klosterneuburg, Austria}
\affiliation{QuEra Computing Inc., 1284 Soldiers Field Road, Boston, MA, 02135, USA}

\author{Sheng-Tao Wang}
\affiliation{QuEra Computing Inc., 1284 Soldiers Field Road, Boston, MA, 02135, USA}

\date{\today}
\begin{abstract}
Combinatorial optimization is a challenging problem applicable in a wide range of fields from logistics to finance. Recently, quantum computing has been used to attempt to solve these problems using a range of algorithms, including parameterized quantum circuits, adiabatic protocols, and quantum annealing. These solutions typically have several challenges: 1) there is little to no performance gain over classical methods, 
2) not all constraints and objectives may be efficiently encoded in the quantum ansatz, and 3) the solution domain of the objective function may not be the same as the bit strings of measurement outcomes. This work presents ``non-native hybrid algorithms" (NNHA): a framework to overcome these challenges by integrating quantum and classical resources with a hybrid approach. By designing non-native quantum variational ansatzes that inherit some but not all problem structure, measurement outcomes from the quantum computer can act as a resource to be used by classical routines to indirectly compute optimal solutions, partially overcoming the challenges of contemporary quantum optimization approaches. These methods are demonstrated using a publicly available neutral-atom quantum computer on two simple problems of Max $k$-Cut and maximum independent set. We find improvements in solution quality when comparing the hybrid algorithm to its ``no quantum" version, a demonstration of a ``comparative advantage".
\end{abstract}
\maketitle

\section{Introduction}

Combinatorial optimization is a challenging task that is widely applicable across a range of domains, from logistics, finance, operations, chemistry, computer science, and more. The underlying problem is deceptively simple: given some problem, such as protein folding or vehicle routing, find a solution in the domain of some objective function that extremizes that function, such as a protein conformation that minimizes energy or delivery route that maximizes utility. Typically, the objective is some low-degree polynomial or other efficient-to-compute function $C(\chi):X\to \mathbb{R}$ that maps some domain of valid solutions $\chi \in X$ to a value capturing the quality of that solution.

\begin{figure}[h!]
    \centering
    \includegraphics[scale=0.85]{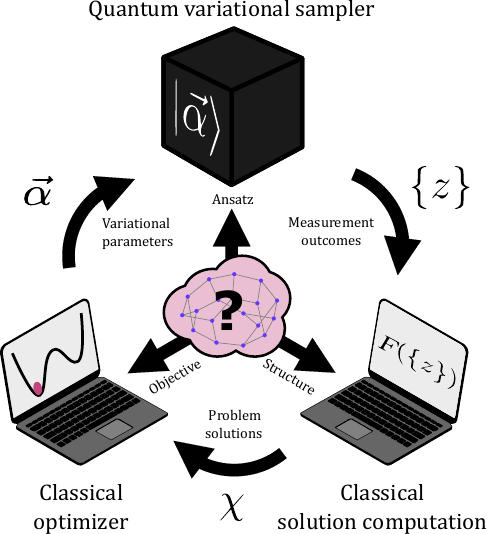}
    \caption{Outline of non-native hybrid algorithms (NNHA). A given problem (center) motivates some problem-dependent ansatz $|\vec \alpha\rangle$. The state is prepared and measured on a quantum computer (top), generating samples $\{z\}$ over the probability distribution $P(z)=|\langle z|\vec\alpha\rangle|^2$. These samples are sent to a classical algorithm $F$ (bottom right), which uses the bit strings to compute a problem solution $F(\{z\})=\chi$, by post-processing individual bit strings, using expectation values as a ``quantum hint", or using the distribution. These classically computed solutions, which do not have to be bit stings, are fed into a classical variational optimizer (bottom left), which optimizes variational parameters $\vec \alpha$ to maximize the objective function $\langle C(\chi)\rangle$. The hybrid algorithm then continues in this loop until convergence.}
    \label{fig:splash}
\end{figure}

These problems are \textit{combinatorial}, in the sense that the domain $X$ of possible solutions is a discrete set given by combinations of subunits, typically growing exponentially in the problem size. This deceptively simple formulation hides extreme complexity; often, general NP-complete problems can be reduced to these optimization problems, which implies that efficient solvers for these problems may collapse the polynomial hierarchy and imply complexity class equivalence $\texttt{P}=\texttt{NP}$. Under the widely accepted assumption that $\texttt{P}\neq\texttt{NP}$, classical optimizers must resort to heuristic ``guess and check" algorithms which search through a large range of candidate solutions and guarantee optimal results only in exponential time.

A promising alternative to classical optimizers is to use quantum computing \cite{abbas2023quantum}, which leverages quantum phenomena to generate approximate or optimal solutions faster or of better quality than existing classical methods. These algorithms fall into two main categories. The first category of \textit{parameterized quantum circuits} \cite{Cerezo2021} uses a digital circuit model parameterized by variational controls $\vec \alpha$ to prepare a state whose probability distribution over measurement outcomes $P(z)=|\langle z|\vec \alpha\rangle|^2$ is biased towards good bit string solutions $z\in X$, maximizing, for example, the average objective function value $\langle C(z)\rangle=\sum_zP(z)C(z)$. Typically, the structure of the problem instance is directly inherited in the parameterized circuit; for example, the quantum approximate optimization algorithm (QAOA) \cite{farhi2014quantum} and its generalizations \cite{blekos2023review} alternate between some global mixing unitary, and a diagonal target unitary generated by accumulating phase proportional to the objective function value $\hat U = \sum_z\exp(-i\gamma C(z)\big)|z\rangle\langle z|$. If the objective function has an efficient low-degree polynomial representation with few hard constraints, such as a quadratic unconstrained binary optimization (QUBO), each clause may be represented by some $k$-local $Z$-phase gate. Alternatively, hardware-native ansatzes \cite{farhi2017quantum} inherit little or no structure from the underlying problem, instead targeting particular hardware-constrained connectivity and optimizing over a larger set of variational parameters.

The second category of \textit{adiabatic quantum computation} \cite{Albash_2018} uses coherent analog dynamics or quantum annealing to prepare the ground state of some physical quantum Hamiltonian. By designing the interaction terms, the ground state directly or indirectly encodes the maxima of the objective function $C(z)$.

Both approaches run into three problems that may otherwise block commercial-scale adaptation, whose problems typically do not fall into clear problem classes. First is that performance over contemporary classical optimization may be small or nonexistent \cite{Marwaha_2021}, with challenging performance guarantees or sampling rates required to beat best-in-class routines \cite{Lykov2023}.

Second is that it may be difficult to encode every constraint and objective directly into the parameterized circuit or Hamiltonian; for example, a problem may have long-range connectivity that is not native to the hardware implementation \cite{Weidenfeller2022scalingofquantum}, include some high-order polynomial term that cannot be efficiently reproduced by a hardware native $k$-local gate set \cite{Hadfield_2021}, or require hard constraints that may be difficult to encode \cite{Hadfield_2019, LaRose_2022}. These incomplete mappings may reduce the performance of the quantum algorithm, as the Hamiltonian ground state may not encode the combinatorial solution, or the parameterized circuit may lose some important context in the problem structure.

Third and most crucially, the domain $X$ of problem solutions may not be the domain of bit strings $\{0,1\}^N$, so even the implementation of a direct encoding \cite{Lucas_2014} may be unclear or suffer from large qubit overhead \cite{Okada2019}. For example, a vehicle routing problem corresponds to an ordered list of locations for a vehicle to travel; while encodings exist \cite{Asad2023},
There is a qubit overhead in mapping from binary outcomes to valid solutions. This challenge is a particular blocker for the utility-scale use of quantum computers solving optimization problems, as realistic instances rarely directly have a binary solution domain.

These challenges are particularly acute when considering real-world optimization problems. These problems are complicated, typically having nontrivial mixed-integer solution spaces with many constraints and multiple objectives that must be optimized in a dynamic or time-constrained manner. Furthermore, these problems typically do not have a natural reduction to specific combinatorial optimization problems, such as MaxCut or SAT, and require some reduction, typically to QUBO at a high overhead cost \cite{Biswas_2017}.

This paper proposes a framework called \textbf{``non-native hybrid algorithms"} (NNHA) that attempts to circumvent these challenges by maximally leveraging both quantum and classical computational resources to solve generic, potentially non-native combinatorial optimization problems, as well as proposing a new metric of ``comparative advantage" over classical algorithms. The key idea, as illustrated in Fig.~\ref{fig:splash}, is to use the quantum device as a \textit{variational sampler} \cite{Amin_2018}, generating bit strings $\{z\}$ from measurement outcomes from some hardware native parameterized ansatz that inherits some but not all structure of the non-native target problem. These bit strings are then used as a resource for a classical algorithm, which uses direct measurement outcomes, correlation functions, or probability distribution to compute some approximate or exact solution $\chi = F(\{z\})$ to the target problem. This contrasts with contemporary quantum optimization, where bit strings directly encode solutions and classical resources only decode bit strings to solutions. In this way, the quantum resources act as a \textit{co-processor} or \textit{hardware accelerator} \cite{cuda} for the overall computation, solving sub-problems or generating biased probability distributions instead of directly solving the target problem.

While the terminology of ``hybrid computing" is well-accepted \cite{DWAVE_hybrid},
this work proposes an extension of the hybrid architectures beyond the paradigm of classical resources simply seeking good variational parameters for some ansatz state \cite{McClean_2016}. Instead, the key idea is that classical resources contribute actively to both parameter finding and directly constructing candidate solutions to the target problem. This work augments other ideas, such as greedy \cite{ayanzadeh2022quantumassisted, Dupont_2023}, recursive \cite{Bravyi_2020, Bravyi_2022,finzgar2023quantuminformed}, multilevel \cite{Ushijima_Mwesigwa_2021, Agone2023}, branch-and-bound \cite{chakrabarti2022universal}, genetic \cite{Schuetz_2022}, pre-processing \cite{Wurtz2021, ponce2023graph}, or linear programming \cite{wagner2023enhancing} approaches.

The rest of the paper is structured as follows. Section \ref{sec:methods} will outline the general structure of the hybrid optimization scheme and various forms (types 1-4) of classical co-processing. Section \ref{sec:maxcut} will demonstrate a type-1 algorithm for the MaxCut problem, which inherits performance guarantees of the classical subroutine as a partial solution to challenge (1) of contemporary quantum optimization. Section \ref{sec:partitioning} will demonstrate a type-2 algorithm for Max $k$-Cut, where the solution domain is not bit strings, a solution to challenge (3). Finally, section \ref{sec:tempering} will demonstrate a type-1 and type-3 algorithm for the maximum independent set problem, where hard constraints cannot be completely imposed by quantum ansatzes, a solution to challenge (2). Examples \ref{sec:partitioning} and \ref{sec:tempering} will also include implementation on QuEra's publicly available neutral-atom quantum computer, Aquila~\cite{wurtz2023aquila}, and are reproducible via code notebooks here \cite{GitHub24}. All examples are intended as prototype demonstrations of the framework instead of general solutions to specific problems.

\section{Methods} \label{sec:methods}

An NNHA algorithm leverages two key components. The first is the quantum variational sampler, which efficiently generates bit strings $\{z\}$ from some probability distribution $P$ parameterized by variational parameters $\vec \alpha$ and problem instance $I$

\begin{equation}
    \{z\}\;\leftarrow\;P(z|\vec\alpha) = \langle z|\hat\rho(\vec\alpha)|z\rangle,
\end{equation}
where $\hat\rho$ is the state generated by the quantum ansatz $\hat \rho\approx |\vec \alpha\rangle\langle \vec\alpha|$ parameterized by problem instance $I$ given a particular algorithm design. Note that there are no strict requirements on coherence or even that the underlying probability distribution be drawn from a quantum device. A quantum device serves as an efficient generator of measurements and may be replaced by e.g.~emulation \cite{Vidal2004}, 
graph neural networks \cite{zhou2021graph}, classical Boltzmann machines \cite{Salakhutdinov2009}, and so forth \cite{Czischek2022}. A ``no-quantum" or classical version of a hybrid algorithm is represented by a state $\hat\rho(\vec\alpha_c)$ that is efficient to sample classically with a set of parameters $\vec\alpha_c$. A natural extension samples from multiple parameterizations $\vec \alpha$, so that each bit string is the concatenation of each measurement outcome in the set of parameterizations. For example, an ansatz may sample individual bit strings from a range of nonequilibrium evolution times, or from several hardware-native best fits of the problem onto a quantum circuit \cite{Hashim2022}.

\subsection{Classical computation}

The second key component of NNHA is classical computation, which serves two main functions. The first is to use bit strings generated from $k$ measurement outcomes into valid problem solutions, through some function

\begin{equation}
    F:\{\{0,1\}^n\}^k\;\mapsto \; X
\end{equation}
which maps ensembles of bit strings to solutions $F(\{z\})=\chi$. The particulars of such a function should be problem-dependent but fall into four main types.

    \noindent \textbf{Type-1.} Computation that \textbf{``post-processes"} individual candidate bit strings to generate individual solutions. For example, the function could use a measurement outcome to directly ``warm start" a classical heuristic optimizer, which refines the initial solution into a locally better solution. This class of hybrid computation is exemplified in Section \ref{sec:maxcut}, which uses a local greedy heuristic to find a local maximum in the energy landscape.
    
    \noindent \textbf{Type-2.} Computation that uses expectation values $\langle \hat z\rangle$ and correlation functions $\langle \hat z_i\hat z_j\rangle$, including higher-order correlation functions, e.g.~$\langle \hat z_i \hat z_j \hat z_k \hat z_l\rangle$, as a \textbf{``quantum hint"}. The values of these expectation values are problem-structure dependent, and shallow circuits may be used to infer the local structure of the problem instance. For example, the connected correlation function can be shown to be strictly local \cite{wurtz_2021_guarantees} for certain ansatzes, and intermediate-depth evolutions can generate correlations within some lightcone that may be larger than a $k$-local classical algorithm can handle. This class of hybrid computation is exemplified in Section \ref{sec:partitioning}, which uses the low-energy eigenvectors of the connected correlation matrix to solve Max $k$-Cut.

    \noindent \textbf{Type-3.} Computation that directly uses the distribution of measurement outcomes $\{z\}\leftarrow P(z|\vec\alpha)$ as a \textbf{``reservoir"}. Certain quantum probability distributions can be hard to sample, which suggests that samples from this distribution may be a natural resource for optimization. For example, if the distribution is known to be approximately thermal, one could use parallel tempering and replica exchange to uniformly sample low-temperature distributions with simulated annealing. An optimization-focused example is exemplified in Section \ref{sec:tempering}, where a maximum independent set is optimized using cluster update simulated annealing.

    \noindent \textbf{Type-4.} While not covered in this work, the notion of \textbf{``iterative refinement"} or recursive schemes can be considered as a NNHA and is the subject of recent research \cite{ayanzadeh2022quantumassisted, Dupont_2023, Bravyi_2020, Bravyi_2022, finzgar2023quantuminformed}. Under a recursive method, measurement outcomes or expectation values are used to fix a subset of problem variables, reducing the problem size. Then, the reduced problem is fed back recursively until the problem size is reduced to zero. While these schemes are ``hybrid", they differ from types 1-3 as they require iterative calls to quantum resources with different ansatzes instead of static distributions $P(z|\vec\alpha)$.

\subsection{Parameter optimization}\label{sec:variational}

The second function of classical computation is variational optimization. Typically, the computational problem can be formulated in terms of a maximization problem for a cost or objective function $C(\{z\})$ \cite{Nannicini_2019}. However, because bit strings are not necessarily valid solutions, the optimizer must search not over \textit{bit strings}, but instead over \textit{classically computed solutions} $C(F(\{z\}))$, so the optimizer finds parameters $\vec\alpha$ that extremize

\begin{equation}
\texttt{MAX}_{\vec\alpha}\;\;\Big\langle C\big(F(\{z\})\big)\Big\rangle \quad\text{with}\quad \{z\}\leftarrow P(z\;|\;\vec \alpha ),
\end{equation}
where the brackets $\langle \cdots\rangle$ denote the average over many samples. The goal of the classical processor is thus to optimize the variational parameters $\vec{\alpha}$ such that the objective function is maximized. This task is typically carried out by a classical optimization algorithm. 

The timescale for operations on a quantum processor is typically orders of magnitude slower than on a classical processor, which motivates the careful choice of an optimization algorithm that is not only resilient to noise but also resourceful and requires few calls to the quantum processor for convergence. We found the Constrained Optimization by Linear Approximations (COBYLA) method to be a suitable choice for this task~\cite{powell1994direct}, though other optimization methods may be appropriate \cite{Bonet2023}. COBYLA does not require the computation of a gradient, which makes it more resilient to hardware noise. The method works by constructing an approximation model of the objective function and constraints in the neighborhood of the current point, which is maximized and updated with each iteration until convergence. 

For each iteration, the cost function value is estimated from a set of $M$ measurement outcomes $\{z\}$. While the variance of the expectation values typically scales as $\sim 1/\sqrt{M}$, we find that optimization with fewer shots per cost estimation can lead to similar or better performance than optimization with many shots for the noise present on neutral-atom quantum hardware. The main driver for good performance is the number of iterations of the optimization algorithm rather than an accurate estimation of the objective function. 
 
While on superconducting hardware $10^3-10^4$ shots are typically used for the estimation of the objective function~\cite{sack2023large-scale}, we found that as few as $\sim10$ shots to work well for optimization on neutral-atom quantum hardware. In the data presented in this work, we typically use $100$ shots. For the variational ansatzes used in this work, we find good convergence within $30$ iterations and $3000$ total shots on the quantum processor. The few shots for convergence may be a feature of analog quantum computation where the number of variational parameters is limited, and the ansatz structure is simple. We expect that for a more general digital ansatz, such as the hardware-efficient ansatz~\cite{peruzzo2014vqe} or unitary coupled cluster ansatz~\cite{mcclean2016theory}, more iterations are required for convergence.

\subsection{Comparative advantage}

A crucial caveat of hybrid algorithms comes in comparing performance between hybrid and classical algorithms. If the hybrid optimization performance is better than some other classical method, is this due to the specific problem subclass, the classical part of the hybrid algorithm, or a clear ``quantum advantage" \cite{Herrmann_2023}? This work uses the insights of designing algorithms with a \textit{classical} or \textit{no-quantum limit} \cite{Wurtz2021}. Given some set of variational parameters, the quantum sampling subroutine may be replaced by an efficient and equivalent classical subroutine, which generates a classical-only variant of the algorithm. For example, the ansatz could be a simple-to-sample product state or matrix product state. In this way, the performance guarantee of the hybrid algorithm is inherited by that of the classical subroutine. The performance of the algorithm is then compared between the classical-only limit, and the full hybrid implementation that does have quantum resources. If the presence of quantum resources increases performance, then there is a comparative advantage to including these resources in computation.

\section{MaxCut with QAOA and greedy FLIP} \label{sec:maxcut}

A simple demonstration of a type-1 NNHA can be shown as applied to the MaxCut problem using an enhancement of the QAOA. For this example, the choice of problem, ansatz, and the classical algorithm is explicitly simple to pedagogically introduce the concept as well as demonstrate the notion of ``comparative advantage". This example can be expanded to more complicated examples as shown in later sections, as well as extended to designing bespoke solvers for application-scale solutions and is intended to serve as an initial inspiration for in-depth bespoke solutions.

MaxCut is an unconstrained binary optimization on graphs. Given a graph $G$ of vertices $V$ and edges $E$, MaxCut strives to find the bipartition of vertices $\{V_+,V_-\}$ such that a maximum number of edges have one vertex in each partition (``cut"). MaxCut is a prototypical NP-hard optimization problem, with classical optimization limited performance guarantees $\leq 0.878$ in the worst case \cite{Goemans1995}.

The QAOA is the prototypical variational quantum optimization algorithm \cite{farhi2014quantum}, which generates an ansatz that alternates $p$ times between some mixer $B$ and target $C$ generator with $2p$ variational parameters $\vec\alpha = \{\vec\beta,\vec\gamma\}$

\begin{equation}\label{eq:qaoa}
    |\vec\alpha\rangle = \prod_{j=1}^p e^{-i\beta_j \hat B}e^{-i\gamma_j \hat C}|+\rangle.
\end{equation}

The target generator $\hat C$ has the property of sharing its eigenspectra with the objective function $\hat C=\sum_z C(z)|z\rangle\langle z|$, and the initial state $|+\rangle$ is the maximal eigenstate of the mixing operator, which is typically taken to be a uniform sum of $X$ terms $\hat B=\sum_j\hat \sigma_x^j$.

\begin{figure}
    \centering
    \includegraphics[scale=1]{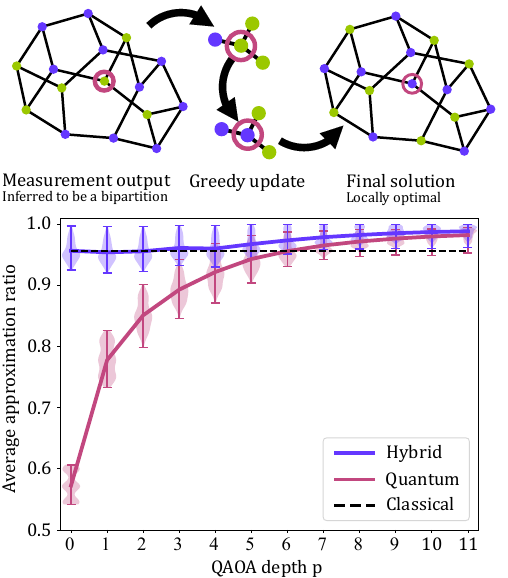}
    \caption{Structure and performance for type-1 NNHA using post-processed QAOA, solving MaxCut on a random ensemble of 256 3-regular 16-vertex graphs. Top shows the action of the greedy flip algorithm. The greedy algorithm iterates through vertices based on some initial bipartition based on quantum measurement (left) and tries to flip partitions based on the 1-local environment, accepting if the objective increases. The process repeats until no further flips can increase the energy, finding a local maximum (right). Bottom: Comparative performance of hybrid, quantum, and classical algorithms. The line is the average expected approximation ratio of each graph over the ensemble, while the violin plots show the distribution of expectation values over the ensemble. Observe that the hybrid greedy post-processing method (purple) outperforms base QAOA (red) and begins to outperform the $p=0$ classical-only limit (black dashed) at $p\sim 4$, indicating an ensemble performance advantage.}
    \label{fig:maxcut}
\end{figure}

MaxCut and its generalization of QUBO is a natural target problem for the QAOA \cite{crooks2018performance,Zhou2020}, due to the simple translation from measurement outcomes to problem solutions. The space of bit string measurement outcomes $\{0,1\}^n$ is in one-to-one correspondence with MaxCut bipartitions, by simply assigning each vertex to the bipartition given by the measurement outcome of the corresponding qubit $0 \mapsto V_+$ and $1\mapsto V_-$. This fact will not be true for later problems.

While QAOA has limited performance guarantees in the small $p$ limit \cite{wurtz_2021_guarantees} and converges to exact solutions in the large $p$ counterdiabatic limit \cite{Wurtz_2022_CD},
The algorithm is typically heuristic and struggles to beat contemporary classical optimizers on a range of problems. A local classical algorithm can strictly outperform $p=2$ QAOA \cite{Marwaha_2021}, and a simple 1-local greedy heuristic can outperform even $p\sim 6$ base QAOA \cite{Lykov2023}, as seen in Fig.~\ref{fig:maxcut}.

QAOA may be enhanced by merging it with a local greedy algorithm using a type-1 NNHA using ``post-processing". The algorithm includes three parts: the ansatz, the variational optimizer, and the classical solution generation. The ansatz is the base QAOA of Eq.~\ref{eq:qaoa}, where the objective generator are the sum of quadratic terms $\hat C = \sum_{\langle ij\rangle \in G} \hat \sigma_z^i\hat \sigma_z^j$ over all edges in the graph $G$, and the mixing generator is the uniform sum over Pauli $X$ terms, $\hat B = \sum_i \hat \sigma_x^i$.
To simplify the analysis, this demonstration uses fixed angles optimized for $\nu$-regular graphs \cite{Wurtz2021_fixed}, bypassing the variational optimization step.

The classical post-processing algorithm is chosen to be a 1-local greedy flip heuristic \cite{hirvonen2014large}, shown in Fig.~\ref{fig:maxcut}. The algorithm is initialized by a single measurement outcome, which is inferred to be an initial candidate solution by mapping a bit string to a bipartition $0 \mapsto V_+$ and $1\mapsto V_-$. Next, the solution is post-processed with the greedy algorithm. Each vertex is trialed to be ``flipped" to be in the opposite partition. Each of these trials increases or decreases the objective value of the solution, depending on the local solution from neighboring vertices. The greedy solver simply picks the vertex flip with the largest increase in objective value; tied vertices are selected at random. The process iterates until no local flip increases the objective value and the algorithm returns a solution. If the algorithm instead chooses a random vertex instead of trialing all vertices, this update rule is the same as zero-temperature simulated annealing; a generalization beyond the scope of this work could include implementation of finite-temperature simulated annealing (e.g. replica exchange); see example \ref{sec:tempering} for more details.

The greedy flip heuristic has a performance guarantee of $\lceil \nu/2\rceil / \nu$ and converges to a solution in a time of order $N^2$ and memory of order $N$ (with a simple generalization to time of order $N$ using zero-temperature simulated annealing \cite{kirk83}). The hybrid algorithm can recover the classical-only limit by choosing all $\beta_i =0$ or doing $p=0$ layers, in which case the sampling probability is uniform $P(z)=2^{-N}$, which is trivial to sample classically. Note that while it may be efficient to simulate expectation values of QAOA using tensor networks or other methods that leverage the strict locality of QAOA, sampling from certain QAOA distributions can be shown to be in the class \#P \cite{farhi2019quantum}, though the reduction does not work for specific instances such as 3-regular MaxCut.

A comparative demonstration of this hybrid algorithm is shown in Fig.~\ref{fig:maxcut} using emulated state-vector simulation. Results are characterized using an ensemble of 256 random 3-regular graphs with 16 vertices. There are a few key takeaways from these results. First, the classical $p=0$ limit of the algorithm outperforms the base QAOA even for depths $p\sim 6$, highlighting the poor performance of quantum-only algorithms against even the simplest classical heuristics.

Second, the shot-to-shot objective value of the hybrid algorithm is strictly greater than or equal to that generated from base QAOA measurements due to the greedy action of post-processing. In this way, any advantage in terms of performance improvements over base QAOA is self-evident. This performance gain is clear when comparing the red and purple lines of Fig.~\ref{fig:maxcut}. While this may be a trivial fact, this notion of maximally using classical resources is the key insight of this work.

Third, for advancing depths of QAOA and thus more quantum resources, the hybrid algorithm heuristically appears to have better performance than both the classical or quantum algorithms individually, as shown by the rise above the black dashed classical-only limit. Unlike the performance increase over quantum-only, this increase is not self-evident and is only a heuristic observation. In principle, the probability distribution over measurement outcomes for the chosen fixed angles may maliciously sample from regions in solution space that have many low-quality local maxima, reducing the performance of the greedy heuristic below the classical-only limit. This phenomenon occurs and is further explored in Sec.~\ref{sec:greedy_mis_ebadi}. It is interesting to explore how optimal parameters may change when optimizing over post-processed objective values, but such a study is beyond the scope of this work.

This simple first example serves only as an initial demonstration of NNHA. There may be many interesting future directions in investigating different problem classes, post-processing methods, approximate or hardware-native ansatzes, variational optimizations, benchmarking against state-of-the-art optimizers, as well as applying this post-processing to real-world optimization problems. Additionally, type-1 post-processing can be used as a subroutine for other NNHA, as will be used in every proceeding example.

\section{Max $k$-Cut with quench dynamics and spectral clustering}\label{sec:partitioning}

\begin{figure*}
    \centering
    \includegraphics[scale=1.0]{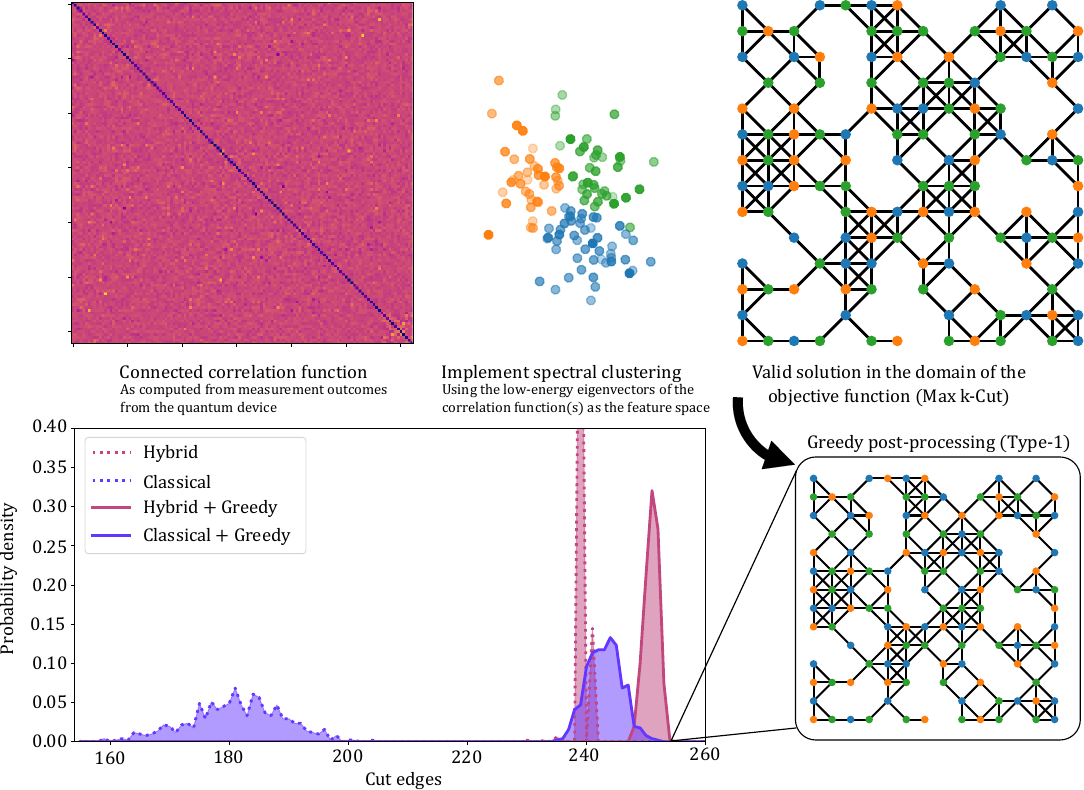}
    \caption{Structure and performance of spectral clustering solving the Max $k$-Cut problem on unit disk graphs, as implemented on QuEra's cloud-accessible quantum computer `Aquila'. Top left plots the variationally optimized connected weighted correlation function (Eq.~\ref{eq:connected_correlation_function}) for the particular problem instance. The three lowest eigenvectors of this matrix serve as a feature vector for $k$-means clustering (top middle) which generates a candidate Max $k$-Cut solution (top right). Each candidate solution can be further post-processed with a zero-local greedy flip algorithm to find a locally optimal solution. The bottom left plots the objective distribution of solutions. The quantum version outperforms the classical-only limit (bottom left) of a greedy post-processed random choice in finding more optimal solutions, which is an indicator of comparative advantage.}
    \label{fig:clustering}
\end{figure*}

An extension of MaxCut is the Max $k$-Cut problem~\cite{Coja2003, buluc2015recent, Bravyi_2022, Agone2023} which seeks to find a $k$ disjoint subsets of a graph such that a maximum number of edges have vertices of different subsets. For $k=2$, the problem reduces to the binary MaxCut problem, and if $k\geq\chi(G)$, the chromatic number of the graph, the optimal solutions are also a $k$-coloring. Due to MaxCut and graph coloring reducing to Max $k$-Cut, which are NP-complete, the decision variant of Max $k$-Cut is also NP-complete.

As discussed in section \ref{sec:maxcut}, $k=2$-MaxCut is a natural problem for quantum optimization since the domain of solutions is directly mapped to bit strings, and the objective function is naturally mapped in $ZZ$ or parity gates. $k\geq 3$ faces the challenge that the direct mapping from bit strings to solutions does not exist. While encoding strategies such as 1-hot exist \cite{Okada2019}, they may become unwieldy for larger $k$ and suffer from limited qubit numbers and coherence.

For these reasons, NNHA may be a viable solution to efficiently extract approximate or exact Max $k$-Cut solutions using non-native quantum hardware. In this section, we propose an example or prototype type-2 NNHA using expectation values and correlation functions that integrate two classical routines: spectral clustering and type-1 greedy post-processing. Keeping with the design intent of NNHA, this implementation will have a natural no-quantum limit and includes both classical solution generation and variational optimization steps. This example implementation should be seen as an example type-2 NNHA, with a specific method and target problem, and should be considered as inspiration instead of generalization to more complicated applications.

Spectral clustering is a method to partition nodes based on eigen-analysis-based features \cite{Ng2001}. The classical algorithm works in the following steps:
\begin{enumerate}
    \item Given a graph $G$, compute the eigenvalues $E_\lambda$ and eigenvectors $V_{i\lambda}$ of the graph Laplacian $J(G)$ or adjacency matrix. The eigenvectors can be interpreted as the ``spring modes" of the graph if every edge can be considered to be a harmonic potential between vertices. The lowest-energy eigenvectors correspond to long-wavelength spring modes with many vertices moving together and the highest-energy eigenvectors correspond to short-wavelength spring modes with adjacent vertices moving opposite.
    \item Identify the largest $k$ eigenvectors $V$ of the Laplacian as a feature vector $F_i$ for each vertex $i$
    \begin{equation}
        \vec F_i = \big( V_{i,N}, V_{i,N-1},\cdots,V_{i,N-k}\big)
    \end{equation}
    If the smallest $k$ eigenvectors are selected, this method will construct an approximate $k$-clustering.
    \item Given the feature vector $\vec F_i$ of each vertex $i$, implement a $k$-clustering algorithm to each vertex to identify each vertex with a color. This implementation uses a simple $k$-means clustering. The output is a vector $S_i$ of colors as a candidate Max $k$-Cut solution.
\end{enumerate}

This solution can be enhanced by type-1 greedy post-processing of the classical output, similar to Sec.~\ref{sec:maxcut}. The algorithm iteratively tries every single-vertex color update, selects the one that decreases the cost function the most, and repeats until no further updates are possible. This algorithm is a generalization over Sec.~\ref{sec:maxcut} to multiple colors.

We can augment this algorithm using a quantum computer with a type-2 NNHA that uses connected correlation functions as a resource for clustering. Instead of using the eigenvectors of the connectivity matrix as the feature vector, we propose to use the eigenvectors of the weighted sum of connected correlation functions

\begin{equation}\label{eq:connected_correlation_function}
C_{ij}(\alpha,\lambda) = \sum_{i=1}^n \lambda_i \big(\langle \vec \alpha_i|\hat n_i \hat n_j|\vec \alpha_i\rangle - \langle \vec \alpha_i|\hat n_i|\vec \alpha_i\rangle\langle \vec \alpha_i|\hat n_j|\vec \alpha_i\rangle\big),
\end{equation}
where $|\vec \alpha_i\rangle$ are $n$ wavefunctions of quantum variational parameters $\vec \alpha$ and classical variational weights $\lambda_i$. There is of course flexibility in the design of ansatz $|\vec \alpha\rangle$; in this case, we use a nonequilibrium quench of the Rydberg neutral-atom Hamiltonian, where each atom is placed at a position scaled from vertex positions on the original graph. For more details, see code notebooks here \cite{GitHub24}. This ansatz requires a constraint on the viable problem instances solved with this algorithm: they must be two-dimensional (2D) geometric with constrained local connectivity, and each vertex must have a position assigned in 2D space.

This work selects problem instances that are King's subgraphs with a 30\% random dropout similar to the graph ensembles of Ref.~\cite{Ebadi_2022}. We select a $k=3$ solution space, as $k=4$ is the prototypical MaxCut problem and any Kings graph can be trivially 4-colored with a tiling.

We choose the variational ansatz $|\vec \alpha_i\rangle$ to be $3$ different quenches from the ground state of the analog neutral-atom Hamiltonian. Each wavefunction is parameterized by $\alpha_i = (t_i,\Delta_i)$, where $t\in[0,4] \, \mu$sec is the total evolution time and $\Delta\in[0,15] \, $rad/$\mu$sec is the detuning. We select a lattice spacing of 4.8$\, \mu$m between sites, and a Rabi drive of $\Omega=15.0\,$rad/$\mu$sec, corresponding to a dynamical blockade radius of $R_b\approx 6.7 \, \mu$m. We find $M=100$ measurements are adequate to recover statistics. Computations are implemented using QuEra computer ``Aquila" through Amazon Braket. For more details, see \cite{wurtz2023aquila} and example notebook \cite{GitHub24}.

Each of the $6$ quantum and $3$ classical variational parameters can be simultaneously optimized. Due to the cost of quantum optimization, this is done dimension-wise: for every set of quantum parameters $\alpha_i$, the classical parameters $\lambda_i$ are optimized using Bayesian methods. Then, the parameter space of quantum parameters is explored using a bounded stochastic search.

Such an ansatz is well physically motivated. At short times and short-range connected quantum systems, correlations and entanglement spread ballistically \cite{Lieb1972} or diffusively \cite{Lux2014} within a small distance. Thus, at times of order $t\sim 1/\Omega$, the connected correlation function between vertex $i$ and others will only be nonzero over a region of radius $\sim 1$ unit cell away. Thus, the eigenvectors should recover some of the local structure of the original graph. By varying the times $t_i$ of each quench, the ansatz probes various local length scales. By varying the weights $\lambda_i$ of the different quenches, one can ``target" a particular length scale. Finally, by varying the detunings $\Delta_i$ of the different quenches, one can infer the local connectivity of each vertex due to the Rydberg blockade effect. In totality, the variational ansatz exploits local correlations and entanglement in the system to extract quasi-local graph structure in a way that may be difficult to reproduce with an equivalent classical analysis.

To recap, the type-2 hybrid spectral clustering algorithm works given the following steps, as shown in Fig.~\ref{fig:clustering}. Given some Max $k$-Cut instance on a unit disk graph $G$ with vertex positions $\vec x_i$, $M$ bit strings are measured from $3$ variational ansatzes of 2 parameters each $\alpha_i$, which is a quench of the Rydberg neutral-atom Hamiltonian. Given $nN$ bit string measurements, the connected correlation function weighted by $n$ classical parameters $\lambda_i$ is computed. The $k$ largest eigenvectors of $C(\alpha,\lambda)$ are used as a feature vector $\vec F_i$ for a $k$-means clustering to identify each vertex with a color, and finally the candidate result is post-processed to be locally optimal with a zero-local greedy flip algorithm. While the expectation values are deterministic (in the $N\to\infty$ measurement limit), candidate solutions are nondeterministic due to the stochastic nature of $k$-means clustering, flip, and finite-sample averaging. Finally, the variational parameters are optimized using Bayesian and stochastic optimizers.

This algorithm, as all NNHA must, has a no-quantum limit. If $t=0$, then the connected correlation function is trivially $C(\alpha,\lambda)=0$, and thus the eigenspectra is degenerate, eigenvectors are Haar random, and the clustering generates a random guess for each vertex color.

Results for this implementation as implemented on QuEra's cloud-accessible quantum computer `Aquila' \cite{wurtz2023aquila} are shown in Fig.~\ref{fig:clustering} for a representative graph and $k=3$. It is clear that the hybrid version of the algorithm outperforms the classical-only limit. This is a clear demonstration of comparative advantage: the presence of quantum resources increases the performance of the algorithm. However, this is not ``quantum advantage", as other structure-aware classical algorithms may be able to beat this heuristic performance.

Finally, it should be noted that this example should not be used directly for solving Max $k$-Cut optimization problems. Instead, it should serve as a \textit{template} for other implementations of type-2 NNHA. Other problems with more complicated objectives and geometries may require other variational ansatzes or classical steps, for which the insight of their design may only come from an understanding of the problem and the action of existing classical optimization routines. Such exploration is beyond the scope of this work.

\section{MIS with quantum annealing and cluster simulated annealing}\label{sec:tempering}

\begin{figure*}
     \centering
     \includegraphics[scale=1.0]{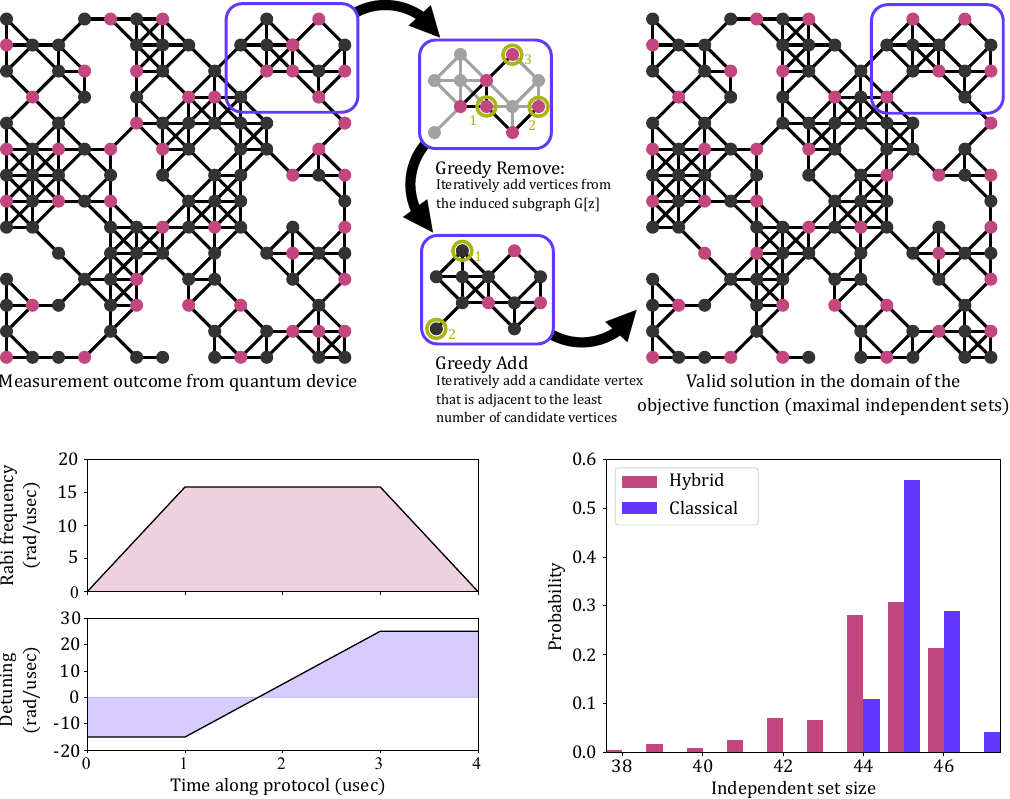}
     \caption{Structure and performance of greedy post-processing solving the maximum independent set on unit disk graphs, as implemented on QuEra's cloud-accessible quantum computer `Aquila'. A measurement outcome from the quantum device (top left) is used as a ``warm start" for a heuristic greedy classical post-processing. First, independent set violations are removed by doing a zero-local greedy add on the induced subgraph $G[z]$, and then the solution is made maximal by randomly adding vertices until no more can be added. The measurement outcomes are generated by a piecewise linear variational adiabatic state preparation (bottom left). Results of the protocol for the example problem instance are shown on the bottom right; for these optimal parameters, the hybrid algorithm is outperformed by the classical algorithm in average solution and maximum solution, indicating no comparative advantage for the particular problem instance. Given a more optimized preparation protocol, it is possible to improve performance and possibly get a quadratic speed-up over classical methods \cite{Ebadi_2022}.}
     \label{fig:greedy_mis}
 \end{figure*}

For the final example, we solve the maximum independent set (MIS) problem using a type-3 NNHA. For completeness, we also discuss the recent results of Ebadi22~\cite{Ebadi_2022}, which can be retroactively interpreted as a hardware demonstration of a type-1 post-processing NNHA, though the encoding itself is very close to native. This example extends the results of that work in a type-3 NNHA by using measurement outcomes as a reservoir for a cluster simulated annealing step.

The MIS problem is a combinatorial graph problem that seeks to find the largest subset of vertices of some graph $G$ such that no two vertices share an edge. The MIS is an NP-complete optimization problem \cite{pichler2018computational} and has worst-case classical performance guarantees that scale logarithmically due to the ``overlap gap" property~\cite{Gamarnik2021}. MIS is a good example of constraint mismatch in quantum algorithms: it is natural to map a bit string $z$ to a set $S=\{i \text{  if } z_i=1\}$. However, not every bit string is a valid independent set (for example, the all-ones bit string), and generating an ansatz with such hard constraints may be difficult. For this reason, some amount of classical processing must happen to get valid solutions. A subset of independent sets are \textit{maximal} independent sets, which are independent sets such that adding any vertex to the set makes the set no longer an independent set. By definition, the maximum independent set must also be a maximal independent set. Certain bit strings (for example, the all-zeros bit string) do not map to valid maximal independent sets. A natural solution space $X$ for MIS is thus the set of all maximal independent sets.

\subsection{Ebadi22 solution}\label{sec:greedy_mis_ebadi}

Recent seminal results by Ebadi22~\cite{Ebadi_2022} solve MIS on a subset of graphs, called unit disk graphs, using neutral-atom quantum computers. The key insight is that the strong Van der Waals interaction between neutral atoms excited to a Rydberg state causes a strong energy shift preventing nearby atoms from both being in the doubly-excited states, in a phenomenon called the ``Rydberg blockade". Given some choice of detuning (e.g.~$Z$ field) and placing atoms at scaled positions of the vertices of the target graph, the Rydberg atom Hamiltonian can encode the maximum independent set of a subset of graphs, called ``unit disk graphs" in its ground state, by matching the Rydberg blockade radius to the unit disk radius~\cite{pichler2018quantum}. The ground state can be prepared adiabatically using protocols such as those of figure \ref{fig:greedy_mis}; for more details see \cite{Ebadi_2022,wurtz2023aquila}.

Although the ground state theoretically encodes the solution, the prepared state may not be the ground state, due to decoherence, SPAM errors, diabatic effects from the ansatz, and more. This is a constraint or objective mismatch between the target problem and the underlying ansatz. For this reason, the measurement outcomes $\{z\}$ may not directly map onto maximal independent sets and require some level of type-1 post-processing. Ebadi22~\cite{Ebadi_2022} used a zero-local greedy heuristic, which works similarly to what follows.

Given some graph $G$ and current independent set $I$, a random vertex is added from the subset of vertices of $G$ that are not in the independent set or adjacent to a member of the independent set. The addition is iterated until no further vertices can be added and the independent set is maximal. This heuristic, which is zero-local as it is not aware of any edge structure when selecting vertices, has a performance guarantee of $2/(\nu+1)$ on $\nu$-regular graphs and is classically optimal up to constant factors \cite{Halldrsson1997}.

The greedy heuristic is used in a two-step process to generate a valid maximal independent set. First, independent set violations are greedily removed as follows. Given some graph $G$ and candidate solution $z$, the greedy add heuristic is implemented starting with the empty set on the subgraph $G[z]$ induced by $z$, generating some independent set $z'$. If $z$ is an independent set, then $G[z]$ has no edges, and thus the independent set $z'$ is the same as $z$. If $z$ is not an independent set, then $z'\subset z$ is an independent set through the greedy add procedure. Likewise, the candidate solution $z'$ may not be a maximal independent set, and thus not a valid solution. By greedy adding warm started by $z'$, the generated solution $\chi$ will be both maximal and independent, and thus be in the solution space. These steps are shown in Fig.~\ref{fig:greedy_mis} Top.

This algorithm has a natural no-quantum limit, by replacing the measurement outcome $z$ with the all-zeros or all-ones bit string. The greedy algorithm will act equivalently on both to add or remove vertices until maximal, and the performance serves as a comparative benchmark against the quantum-enhanced version.

Results for this algorithm as implemented on QuEra's cloud-accessible quantum computer `Aquila' \cite{wurtz2023aquila} are shown in Fig.~\ref{fig:greedy_mis} bottom right, which plots the probability distribution of maximal independent set sizes for the particular graph instance shown; we find that the hybrid  (classical) algorithm generates independent sets of average size 44.282 (45.178) with the hybrid algorithm having a performance ratio of 0.9801, indicating that the hybrid algorithm is outperformed on average by the classical-only limit. Similarly, the classical-only algorithm finds a better single-shot solution than quantum with an 81.2\% probability and finds a MIS of size 47 with a 1.80\% probability, which the hybrid algorithm never finds. These results suggest that, in this case, the quantum elements of the algorithm in fact \textit{hinder}, not \textit{help} the performance. This is a crucial fact: including quantum resources in an NNHA does not result in a self-evident improvement in performance over classical-only algorithms. Indeed, any algorithm must be carefully designed to avoid reducing performance. For example, increasing the complexity and length of the adiabatic protocol may lead to improved performance, which is the case for \cite{Ebadi_2022}.

\begin{figure*}
    \centering
    \includegraphics[scale=1.0]{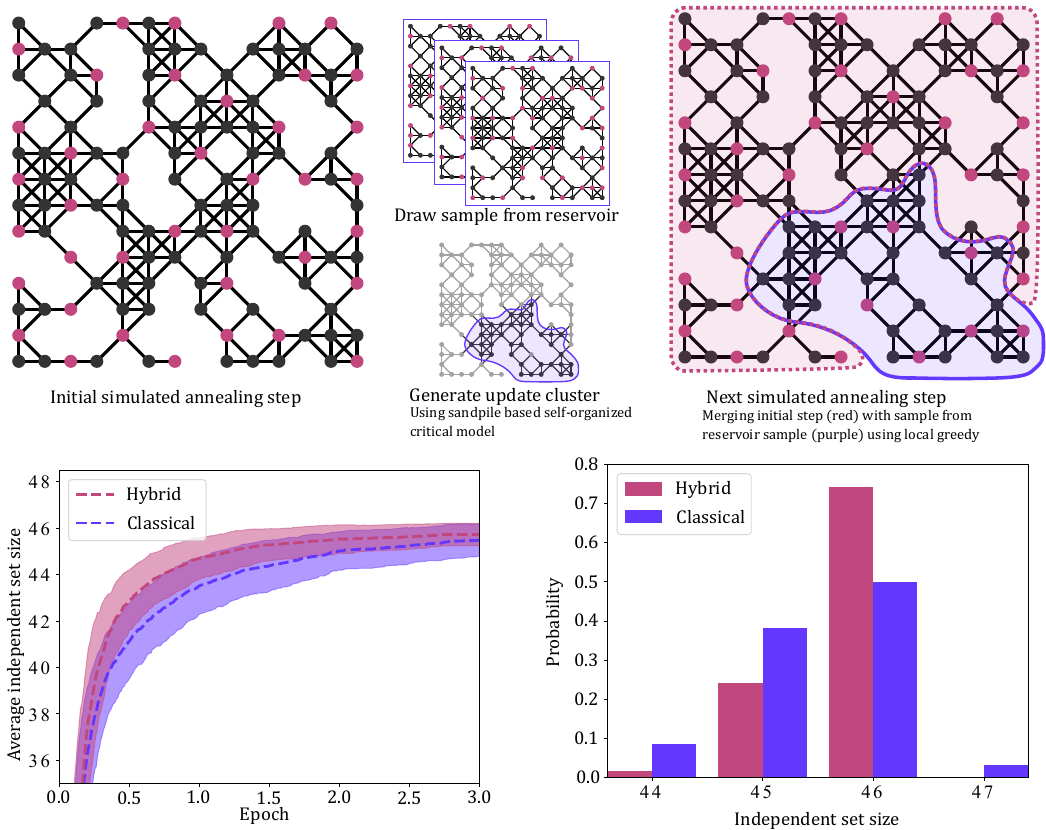}
    \caption{Structure and performance of cluster update simulated annealing implementation using type-3 NNHA, as implemented on QuEra's cloud-accessible quantum computer `Aquila'. Top sketches the structure of the algorithm, which iteratively updates and improves a candidate solution (top left). First, a sample is drawn based on a measurement outcome from a neutral-atom quantum annealing protocol, and a cluster is generated based on a self-organized critical sandpile model (top middle). Then, the candidate solution is merged with the new sample (top right) by replacing the solution within the cluster with the reservoir measurement. The process is iterated many times until converged (bottom left), where each epoch is $N$ update steps. Given variationally optimized parameters, it is clear that the NNHA converges to optimal solutions faster than the classical-only limit (purple). This is an indicator of comparative advantage, as the algorithm performs better when quantum resources are included. The bottom right plots the distribution of solutions after 3 epochs of zero-temperature updates. While the NNHA has a better average performance, the classical variant can occasionally outperform and find the actual MIS of size 47.}
    \label{fig:cluster_update}
\end{figure*}

\subsection{Cluster update: type-3 solution}

A natural example of type-3 NNHA is cluster update using parallel tempering \cite{Geyer1991MarkovCM} and simulated annealing \cite{kirk83,cain2023quantum}, where subsystems are exchanged between two Boltzmann distributions of solutions at different temperatures. By coupling the low-temperature ensemble to the high-temperature one, a Markov update Monte Carlo can be kicked out of local maxima to more efficiently explore the low-energy state space of the distribution. In this implementation, instead of two classically generated distributions, the high-temperature distribution is replaced by samples drawn from the quantum device. The distribution may not be thermal with respect to the objective function, though recent work on QAOA finds that Boltzmann distributions are surprisingly common \cite{Lotshaw2023}. If the ansatz adiabatically prepares almost-ground states of a Hamiltonian that almost encodes the objective function, as is the case for MIS and neutral atoms, the direct samples from the reservoir may nonetheless be close to the maximum, and thus the ensemble may generate relatively low-temperature states~\cite{Wild2021}.

Instead of a direct parallel tempering implementation which occasionally directly exchanges samples from different reservoirs, a more advanced version instead exchanges subsystems of samples from each reservoir, e.g.~a ``cluster update". In this case, only a small portion of the current Monte Carlo solution step in the low-temperature reservoir is replaced with the high-temperature distribution. Cluster update methods with scale-free sizes have been found to converge to better results faster than single-vertex updates \cite{Hoffmann2018}. Note that because many individual quantum samples are aggregated into a single solution, the resulting output may be better than any individual sample.

The cluster update NNHA works as follows, as shown in Fig.~\ref{fig:cluster_update}. Given some solution $\chi$ (initialized as the empty independent set),
\\
\noindent \textbf{1.} Pick some subsystem to update. There are several strategies on how to select a cluster. For instance, one could select a pair of adjacent vertices for which one is in the independent set, in which case the method reduces to a diffusive search over Hamming distance 2-adjacent maximal independent sets \cite{cain2023quantum}. Alternatively, one could select vertices within a fixed radius of a central vertex, in which case the method diffusively explores across larger Hamming distance moves. In this example, the insights of self-organized criticality for optimization are used by selecting clusters based on a sandpile model \cite{Hoffmann2018}. A sandpile model works by iteratively adding ``sand grains" to random vertices (slow process) until a vertex reaches a collapse threshold given by its adjacency; then the vertex sheds a grain to each adjacent vertex (fast process), which may trigger further collapses. This fast process is iterated until no further vertices collapse; the cluster is generated by all vertices that underwent collapse. A non-scale-free dissipation term is added by removing each grain from a collapsing vertex with a 5\% probability. This process is known to generate a scale-free size distribution of clusters. It has been found on similar objectives that such a cluster update outperforms fixed-cluster optimization techniques \cite{Hoffmann2018}.
\\
\noindent \textbf{2.} Sample an approximate solution $\chi$ from the high-temperature reservoir by calling the greedy algorithm of Sec.~\ref{sec:greedy_mis_ebadi} as a subroutine. The no-quantum limit of that subroutine serves as the no-quantum limit of this algorithm.

\noindent \textbf{3.} Merge the solution $\chi$ with the reservoir solution $\chi'$, as $z=\chi\cup \chi'$. In the vertices of the cluster selected by step (1), update the solution $\chi$ to the solution of $\chi'$; outside, keep the solution $\chi$.
\\
\noindent \textbf{4.} Enforce that the new candidate solution $z$ is a valid maximal independent set with a type-1 post-processing, by a mapping $F(z)\to \chi$ using the greedy heuristic of \ref{sec:greedy_mis_ebadi} to guarantee the new solution is a valid maximal independent set. Even if both $\chi$ and $\chi'$ are valid maximal independent sets, the merged candidate may have independent set violations along the boundary.
\\
\noindent \textbf{5.} Accept or reject the update based on the change in objective value using the Metropolis-Hastings update rule \cite{Hastings1970} $P_\text{accept}=\text{MIN}\big[1,\exp(-\beta \Delta)\big]$, where $\beta$ is the inverse temperature and $\Delta = C(\chi') - C(X)$ is the change in objective value.
\\
\noindent \textbf{6.} Finally, go to (1) for some fixed number of steps, or break after the solution $\chi$ becomes stationary given $\beta\to\infty$. Update steps equal to the number of vertices in the graph is one epoch.

Results for this algorithm as implemented on QuEra's cloud-accessible quantum computer `Aquila' \cite{wurtz2023aquila} are shown in Fig.~\ref{fig:cluster_update}, which is implemented for a particular graph instance. Parameters are variationally optimized to an objective function that averages the objective over 3 epochs of annealing steps. A bounded stochastic optimizer finds the best parameters to be $t_\text{max}=3.80 \, \mu$s, $\Delta_{\text{min}}=-13.47 \,$rad/$\mu$sec, and $\Delta_\text{max}=41.95 \, $rad/$\mu$sec. We find that the hybrid (classical) algorithm generates independent sets of size 45.72 (45.48) with the hybrid algorithm having a performance ratio of $1.005$, indicating that the NNHA slightly outperforms the classical-only limit. Similarly, the NNHA samples an equal-or-better solution with an $83\%$ probability. For more details, see an example code notebook here~\cite{GitHub24}.

For this particular instance, the hybrid annealing algorithm converges to finding larger independent sets faster than the equivalent classical algorithm, which simply greedily adds vertices within the reservoir. This may be as expected, as the independent set samples ``know" more about the global structure of the graph than the zero-local greedy algorithm. We observe that performance is highly dependent on the variational parameters. For example, if the final detuning is too small, the effective blockade radius is too large, and the Rydberg states will typically be further apart. This causes measurement samples to accidentally block good maximal solutions, and the NNHA converges to solutions \textit{slower} than the classical-only limit. Such behavior may be useful if, for example, the solution space is over minimum dominating sets, but is otherwise detrimental.

Furthermore, even though the quantum variant finds a larger expected independent set size on average, the classical-only limit may occasionally find a larger independent set size of 47. While this result is seen for a specific problem instance, this still suggests that classical algorithms may outperform NNHA depending on the particulars of the objective. Improvements to this prototype implementation may also see an enhanced performance, though such a study is beyond the scope of this work.

Finally, it should be emphasized that this particular implementation is a \textit{template} for other type-3 postprocessing implementations. For example, other kinds of cluster update strategies may be used, or optimized over different objectives such as dominating sets. In particular, cluster updates may be used to solve MIS problem instances much larger than can fit on quantum hardware. By implementing quantum annealing dynamics on each (subextensive) cluster plus some fixed radius around a cluster, a large graph can be broken into many small subgraphs and stitched together using annealing. In this way, arbitrary large graph problems may be solved using a reasonably sized neutral-atom quantum computer, at the cost that correlations in each sample will be small in extent. Such an implementation is beyond the scope of this work. Alternatively, the samples could be used to generate clusters instead of the updates themselves in a multilevel algorithm \cite{Agone2023}.

\section{Conclusion and future directions}\label{sec:conclusion}

There are three key challenges to contemporary quantum optimization. First, the performance of heuristic quantum algorithms is typically quite poor in comparison to even simple contemporary classical optimization algorithms. Second, the constraints and objectives may not be native to the underlying hardware, requiring expensive overhead and encoding. Finally, the solution domain of relevant real-world problems is not typically bit strings, typically requiring some encoding overhead. 

These challenges are highlighted when attempting to solve real-world optimization problems with quantum or hybrid methods. Typical problems are large, complicated, multi-objective, constrained, and do not naturally or natively map onto quantum hardware.

This work has presented a framework of ``non-native hybrid algorithms" (NNHA) as a method to overcome these challenges, by incorporating quantum resources to assist classical optimization so that the quantum execution does not need to directly encode solutions. In this way, the quantum computer acts as a \textit{coprocessor} for classical computation, offloading specific sub-parts of the optimization while the classical components refine and include more complex real-world components of the problem instance. Such an architecture is particularly natural for high-performance computing applications, where classical computing resources may be much larger than any quantum resource in the near future. Notably, this version of ``hybrid" computing goes beyond the paradigm of only using classical resources for variational optimization, by also directly assisting with computing solutions.

The major contribution of this work is to propose the idea of measurement outcomes from the quantum device $\{z\}$ not needing to directly map to solutions. Instead, NNHA uses bit strings indirectly as a \textit{resource} for a classical step to compute a solution. In this way, classical resources contribute actively to both parameter finding and directly constructing candidate solutions to the target problem. This contrasts with contemporary quantum optimization, where bit strings directly encode solutions and classical post-processing only decodes bit strings into solutions \cite{blekos2023review}. We enumerate with examples three kinds of NNHA: type 1 uses individual bit strings as a ``warm start" to post-process and refine solutions; type 2 uses expectation values and correlation functions as a ``quantum hint" to glean hard-to-find structure of problem instances; and type 3 directly uses the distribution of bit strings as a ``reservoir" for computation.

While this work presents prototype solutions to demonstrate each type of hybrid optimization, they are by no means complete. Indeed, these implementations should be seen as \textit{templates} and inspiration for bespoke solutions to real-world optimization applications. Instead, these methods should serve the key objective of solving optimization problems beyond simple problem classes and including the full complexity of real-world objectives. By maximally leveraging classical resources and designing NNHA that closely synergizes both quantum and classical subroutines, we come closer to a goal of the field: quantum practicality, where quantum resources improve outcomes and solve problems in a practical, demonstrable, and real-world setting.

\section*{Acknowledgements}

This work was supported by the DARPA ONISQ program (Grant No.~W911NF2010021) and DARPA-STTR award (Award No.~140D0422C0035). The authors thank Alexander Keesling, Maddie Cain, Nate Gemelke, and Phillip Weinberg for helpful discussions, and Danylo Lykov, who had early contributions to this work.

\bibliographystyle{apsrev4-2}
\bibliography{qis}
\end{document}